# HABITABLE CLIMATE SCENARIOS FOR PROXIMA CENTAURI B WITH A DYNAMIC OCEAN

Anthony D. Del Genio[1], Michael J. Way[1], David S. Amundsen[1,2],
Igor Aleinov[1,3], Maxwell Kelley[1,4], Nancy Y. Kiang[1],
and Thomas L. Clune[5]

[1]NASA Goddard Institute for Space Studies, 2880 Broadway, New York, NY 10025

[2]Department of Applied Physics and Applied Mathematics, Columbia University, New York, NY 10027

[3] Center for Climate Systems Research, Columbia University, New York, NY 10027

[4]Trinnovim, LLC, 2880 Broadway, New York, NY 10025

[5]NASA Goddard Space Flight Center, Greenbelt, MD 20771

Corresponding author:
Anthony D. Del Genio
NASA Goddard Institute for Space Studies
2880 Broadway, New York, NY 10025
Phone: 212-678-5588; Fax: 212-678-5552
Email: anthony.d.delgenio@nasa.gov

Running title: Habitability of Proxima Centauri b

**ABSTRACT**

The nearby exoplanet Proxima Centauri b will be a prime future target for characterization, despite questions about its retention of water. Climate models with static oceans suggest that an Earth-like Proxima b could harbor a small dayside region of surface liquid water at fairly warm temperatures despite its weak instellation. We present the first 3-dimensional climate simulations of Proxima b with a dynamic ocean. We find that an ocean-covered Proxima b could have a much broader area of surface liquid water but at much colder temperatures than previously suggested, due to ocean heat transport and depression of the freezing point by salinity. Elevated greenhouse gas concentrations do not necessarily produce more open ocean area because of possible dynamic regime transitions. For an evolutionary path leading to a highly saline present ocean, Proxima b could conceivably be an inhabited, mostly open ocean planet dominated by halophilic life. For an ocean planet in 3:2 spin-orbit resonance, a permanent tropical waterbelt exists for moderate eccentricity. Simulations of Proxima Centauri b may also be a model for the habitability of planets receiving similar instellation from slightly cooler or warmer stars, e.g., in the TRAPPIST-1, LHS 1140, GJ 273, and GJ 3293 systems.

Keywords: Proxima Centauri b, habitability, climate, exoplanets, aquaplanet, general circulation model

## 1. INTRODUCTION

The discovery of the exoplanet Proxima Centauri b orbiting the star nearest to Earth (Anglada-Escudé et al., 2016) ranks as one of the most exciting developments in the history of exoplanet science. Proxima b's minimum mass of 1.27 +0.19/-0.17 Earth masses ($M_{\oplus}$) is in the range for possible rocky planets. Its equilibrium radiating temperature of 220-240 K and incident stellar flux (~0.65 times that for Earth) are within the ranges traditionally used to identify targets for analysis of potential habitability (Kopparapu et al., 2013). Its proximity to Earth will provide opportunities for it to be further observed and possibly characterized in the coming years, even though transits are unlikely (Kipping et al., 2017).

On the other hand, Proxima b closely orbits a red dwarf star. This would probably have driven it into a runaway greenhouse state during the star's luminous pre-main sequence phase, and even today would subject it to X-ray and extreme ultraviolet fluxes and stellar winds leading to catastrophic atmospheric and surface water loss (Ribas et al., 2016; Airapetian et al., 2017; Barnes et al., 2017; Dong et al, 2017). Nonetheless, Proxima b has a poorly constrained and potentially complex evolutionary history. Depending on its initial water inventory, the presence of an initial shielding hydrogen envelope, velocities of prior impacts, and whether it migrated to its current position from a greater distance, scenarios exist by which the planet could be habitable (Ribas et al., 2016; Barnes et al., 2017; Zahnle and Catling, 2017).

Thus, it is worthwhile to use the few existing constraints to assess whether hypothetical atmospheres for Proxima b are conducive to life. Meadows et al. (2017) conducted 1-D simulations of photochemically consistent climates for Proxima b that depend on its history. Turbet et al. (2016) used a 3-dimensional (3-D) general circulation model (GCM) with a dry land surface, ice surface, or static thermodynamic ocean. They found "eyeball Earth" climates (Pierrehumbert, 2011) with a substellar region of warm liquid water for an atmosphere similar to modern Earth's if Proxima b synchronously rotates, but required elevated $CO_2$ to maintain an equatorial waterbelt if the planet is in 3:2 spin-orbit resonance. Boutle et al. (2017) used a different GCM with an atmosphere bounded by a 2.4-m liquid surface with fixed albedo and irradiated by a slightly lower stellar flux than that used by Turbet et al. (2016). [The 881.7 $Wm^{-2}$ flux used by Boutle et al. (2017) matches the best estimate of Anglada-Escudé et al. (2016), but the 956 $Wm^{-2}$ value used by Turbet et al. (2016) is within the uncertainties.] Their model produces similar climates to those of Turbet et al. (2016) but with cooler dayside and much colder nightside temperatures. Boutle et al. (2017) find that if spin-orbit resonance occurs with significant eccentricity, two eyeball regions of surface liquid water exist on opposite sides of the planet, the sides that face the star at successive periastrons (Dobrovolskis, 2007; Brown et al., 2014).

The assumption of a static ocean is computationally convenient (and is thus the standard for GCM exoplanet studies) but does not capture much of the physics by which oceans modulate climate. Hu and Yang (2014) coupled a dynamic ocean to an exoplanet GCM to show that ocean heat transport broadens the region of surface liquid water. Yang et al. (2014) used dynamic ocean, sea ice, and land ice models to explore water trapping on the nightside of a synchronously rotating planet. Cullum et al. (2016) used an ocean GCM to emphasize the role that varying salinity might play in determining exoplanet climates.

In this paper we present the first simulations of potential Proxima Centauri b climates using an atmospheric GCM coupled to a dynamic ocean GCM. We consider shallow all-ocean "aquaplanets" and planets with an Earth-like land-ocean distribution. Section 2 describes our model and experiments, and the resulting climates are analyzed in Section 3. The implications for Proxima Centauri b, and for exoplanet habitability in general, are discussed in Section 4.

## 2. METHODS

*a.) Model description*

We use the ROCKE-3D (Resolving Orbital and Climate Keys of Earth and Extraterrestrial Environments with Dynamics) GCM (Way et al., 2017a), a planetary adaptation of the NASA Goddard Institute for Space Studies Model E2

Earth GCM. ROCKE-3D has been used to simulate a potential habitable ancient Venus (Way et al., 2016), the climate response to long-term evolution of planetary eccentricity (Way et al., 2017b), factors controlling the onset of moist greenhouse stratospheric conditions on aquaplanets (Fujii et al., 2017), and possible snowball periods in Earth's past (Sohl et al., 2017).

For the experiments described here, ROCKE-3D was run with an atmosphere at 4°x5° latitude-longitude resolution and 40 vertical layers with a top at 0.1 hPa. This resolution should be adequate given the rotation period of Proxima Centauri b reported by Anglada-Escudé et al. (2016) (11.2 d for synchronous rotation, 7.5 d for 3:2 spin-orbit resonance), which places it near or beyond the threshold at which baroclinic eddies grow to planetary size and the atmospheric dynamics transitions from a rapidly rotating to a slowly rotating regime (Del Genio and Suozzo, 1987; Edson et al., 2011).

We use the SOCRATES[1] radiation parameterization (Edwards, 1996; Edwards and Slingo, 1996) option (identical to that used by Boutle et al., 2017) as described in Way et al. (2017a). SOCRATES employs a two-stream approximation with opacities treated using the correlated-k method based on the HITRAN2012 line list (Rothman et al., 2013). Overlapping absorption is treated using adaptive equivalent extinction (Edwards, 1996; Amundsen et al., 2017). For water vapor, the CAVIAR continuum (Ptashnik et al., 2011) is employed.

---

[1] https://code.metoffice.gov.uk/trac/socrates

Campargue et al. (2016) suggest that CAVIAR overestimates near-infrared water vapor self-continuum absorption, but Fujii et al. (2017) find that differences between CAVIAR and the more weakly absorbing CKD continuum produce very small effects on the climate of planets orbiting M stars in ROCKE-3D. The baseline shortwave (SW) component of SOCRATES used for most of our simulations has 6 spectral bands and the longwave (LW) component 9 bands (Tables 1, 2 of Way et al., 2017a), producing fluxes accurate to within several Wm$^{-2}$ for weakly irradiated planets orbiting M stars with compositions similar to modern Earth. For simulations with elevated $CO_2$ and $CH_4$ concentrations we instead employ 29 SW and 12 LW bands (Table 1) to achieve similar flux accuracy for Proxima b. ROCKE-3D also employs a mass flux cumulus parameterization with two updraft plumes of differing entrainment rates, convective downdrafts, and interactive lifting and detrainment of convective condensate into anvil clouds; a stratiform cloud scheme with prognostic condensate, diagnostic cloud fraction, and cloud optical properties based on simulated liquid and ice water contents and a temperature-dependent ice crystal effective size (Edwards et al., 2007); and a dry turbulent boundary layer scheme with non-local turbulence effects, all described further in Way et al. (2017a).

Tian and Ida (2015) find that for low mass stars, ocean planets and desert planets are more likely final states for habitable zone rocky planets than planets with partially exposed land. Nonetheless, given Proxima b's unknown history, we consider several possible scenarios. For most simulations, the atmosphere is bounded by a dynamic aquaplanet ocean having the same horizontal resolution as the atmosphere and 9 layers extending to 900 m depth, much shallower than the ~3700 m mean depth of Earth's oceans. This is primarily done for computational convenience – even with a fairly shallow ocean, our experiments can require ~1000-5000 orbits (~30-150 Earth years) to reach equilibrium. Nonetheless, even at 900 m depth, the ocean heat capacity (which is proportional to depth) implies a thermal response time to radiative forcing that is much longer than the rotation period of Proxima b. This is important for asynchronously rotating scenarios.

Transport by unresolved ocean eddies is parameterized by a unified Redi/Gent-McWilliams scheme (Redi, 1982; Gent and McWilliams, 1990; Gent et al., 1995). It is not known how applicable such schemes are for oceans on slowly rotating planets, where the spatial scale of the relevant eddies should differ from those on Earth. However, a sensitivity test that replaces the nominal 1200 $m^2s^{-1}$ diffusivity with a much lower value (100 $m^2s^{-1}$) changes maximum and minimum surface temperatures by no more than half a degree.

Ocean albedo is spectrally dependent and varies with solar zenith angle and wind speed. Typical substellar values for the diffuse/direct components are 0.072/0.039 for wavelengths $\lambda < 0.770$ μm and 0.056/0.032 at longer $\lambda$, respectively. Absorption beneath the ocean surface is parameterized with a red/near-IR component with e-folding penetration depth of 1.5 m and a shorter wavelength component with e-folding depth of 20 m, weighted by the solar spectrum for an average turbidity of seawater. To test the sensitivity to the spectrum we performed a variation on *Control* attenuating the entire Proxima Centauri spectrum with the 1.5 m e-folding depth, since 99.4% of the star's SW is in the red/near-IR vs. 40% for the Sun. Because of the small amount of starlight reaching the ocean surface and the 12 m thickness of our first ocean layer, global mean surface temperature is within a degree of *Control*. Local deviations can be larger but only on the nightside. Subsurface temperatures are within 0.3°C almost everywhere. Local increases/decreases of sea ice cover of a few percent occur along the sea ice margin. Despite this, the impact on life underwater could be dramatic. The photic zone where photosynthetic life could be supported, assuming useful radiation at 400-1100 nm and a photon flux cutoff similar to the low-light limit in the Black Sea (Overmann et al., 1992), is effectively < 10 m in this experiment vs. 80 m on Earth, and even shallower at higher zenith angle.

The ocean model includes a dynamic sea ice parameterization (see Way et al., 2017a) with albedos in one visible and five near-infrared spectral intervals that vary with ice thickness, the presence, depth, age, and type (dry or wet) of snow, the presence of melt ponds, and zenith angle. Seawater freezes at a temperature that depends on salinity. Fractional gridbox sea ice cover is permitted.

For simulations with an Earth land-ocean distribution, the land surface is assumed to have an albedo of 0.2, typical of a terrestrial desert. The soil texture is assumed to be a mixture with sand/silt/clay fractions of 0.5/0.0/0.5 (as in Way et al., 2016), giving a porosity of 0.4855 pore volume over soil volume. The GISS land surface model is limited to a depth of 3.5 m, so with this soil texture, the maximum soil water content is 1699 kg m$^{-2}$. Given the lack of seasonality in our simulations, the soil moisture memory that could arise from different soil textures is not an influence on our equilibrium climate. Land areas can accumulate snow to grow ice sheets but land ice dynamics are not included.  The ocean in these partial exposed land experiments is assumed to have flat bathymetry, with no islands.

Planetary properties for Proxima b are identical to those used by Turbet et al. (2016): gravity = 10.98 ms$^{-2}$ and radius = 7127.335 km, based on assumptions that the planet mass $M_p$ = 1.4 $M_{\oplus}$ and the planet density $\rho = \rho_{\oplus}$ = 5.513 g cm$^{-3}$. We use the Proxima Centauri stellar spectrum created by the Virtual Planetary Laboratory and described by Meadows et al. (2017).

*b.) Experiments*

Table 2 describes our simulations. Our baseline experiment, *Control*, assumes a 0.984 bar atmosphere (Earth's mean surface pressure), primarily $N_2$ but with 376 ppmv $CO_2$, an aquaplanet dynamic ocean, synchronous rotation (11.2 d period), and the instellation value of Boutle et al. (2017). Salinity is initialized at a typical Earth value of 35.4 psu (practical salinity units, equivalent to g of salt per kg of water, or 35.4 g $L^{-1}$, or 3.54%). Other experiments use *Control* as the baseline unless otherwise specified. To isolate the effect of the dynamic ocean, we perform a simulation with a 100 m thermodynamic ocean with no ocean heat transport (*Thermo)*. *Control-High* uses the higher instellation value assumed by Turbet et al. (2016). Three *Archean* simulations assume a 0.984 bar $N_2$-dominated atmosphere with elevated greenhouse gas concentrations based on previous studies of Archean Earth. *Archean Med* assumes the $CO_2$ and $CH_4$ abundances of Case A of Charnay et al. (2013). *Archean Low* assumes $CO_2$ and $CH_4$ levels half way between those of *Control* and *Archean Med. Archean High* has $CO_2$ and $CH_4$ concentrations identical to Charnay et al. (2013) Case B.

Two experiments test the effect of ocean salinity (S), based on the study of Cullum et al. (2016): S = 0 psu (*Zero Salinity*), and S = 230 psu (*High Salinity*), a value almost as high as that of the Dead Sea and in the range for extreme halophiles (Ollivier et al., 1994; Oren, 2002). The 230 psu value is slightly smaller than that assumed by Cullum et al. (2016) since our simulations are for a colder planet; it is slightly smaller than the 233 psu value at the eutectic point for an $NaCl$-$H_2O$ mixture, ensuring that our ocean remains below salt saturation. We explore climates for a 3:2 spin-orbit resonance state in simulations with orbital eccentricity e = 0 (*3:2e0*) and e = .30 (*3:2e30)*. Two experiments assume an Earth-like land-ocean distribution, one with the substellar point over the dateline and thus dayside heating of a mostly enclosed ocean basin (*Day-Ocean*), and another with the substellar point at 30°E longitude and thus dayside heating of Africa and parts of Eurasia (*Day-Land*).

## 3. RESULTS

Table 3 lists relevant climate variables for each simulation. All except one produce a regionally habitable surface climate as judged by the presence of areas with time mean temperature above the freezing point of seawater. Despite this the planetary mean surface temperature for each simulation is below freezing, in some cases by tens of degrees. Thus, judged by traditional 1-D habitable zone criteria, none of the planets would be considered habitable. Maximum surface temperatures in the substellar region differ by 34°C among the simulations, due to factors other than instellation or atmospheric composition differences. The global sea ice (or sea ice + snow for simulations with land) cover ranges from 13% to 80%. Minimum temperatures vary more (by 50°C), reflecting the control of nightside temperatures by heat and moisture transports (Yang and Abbot, 2014), which vary considerably (by design) among the experiments.

*a.) Synchronously rotating aquaplanets with an Earth-like ocean and atmosphere*

Figures 1a and 2a show the surface temperature and sea ice cover, respectively, for *Control.* These differ greatly from the "eyeball" pattern of substellar liquid water found by Turbet et al. (2016) and Boutle et al. (2017).

*Control* instead produces a "lobster" pattern of open ocean water that is characteristic of simulations of synchronously rotating planets with a dynamic ocean. As explained by Hu and Yang (2014), this is a realization of the Matsuno (1966) and Gill (1980) pattern of atmospheric response to a stationary heat source, consisting of a Rossby wave-induced cyclonic circulation on either side of the equator west of the heat source and an equator-straddling Kelvin wave east of the heat source. The net result of these two wave modes is a surface wind pattern that converges near the instellation maximum (Fig. 3a), leading to rising motion there and the onset of moist convection and thick clouds (see Section 4b) that shield the surface from much of the incident starlight (Yang et al., 2013).

The near-surface ocean in *Control* develops a mostly westerly equatorial current (Fig. 3b) that advects warm water downstream and keeps the equatorial ocean at least partly ice-free all the way to the anti-stellar point and beyond, similar to that seen in Hu and Yang (2014). Upstream of the instellation peak, strong advection of sea ice into the substellar region (Fig. 3c) and subsequent melting keep equatorial temperatures slightly cooler than on either side of the equator. The ocean is completely ice-free only in a tropical region centered approximately on the evening terminator (Fig. 3d). Because of the night-to-day sea ice mass flux, sea ice thickness on the nightside remains below 10 m (Fig. 3d). The net result of the atmospheric, oceanic, and sea ice transports is a peak surface temperature on the dayside (3°C) that is much cooler than the ~27°C found by Turbet et al. (2016) for the same atmospheric composition. The habitable fraction of the surface $f_{hab}$, defined here as the fraction without sea ice (or sea ice + snow, for the experiments with land described later), is .42.

Figures 1b and 2b show surface temperature and sea ice, respectively, for the static ocean case *Thermo*. The maximum temperature (19°C) is much warmer than that of *Control* despite an identical atmospheric composition, and its minimum is even more dramatically colder, due to the absence of ocean heat transport. Our dayside maximum is comparable to that of Boutle et al. (2017) but ~10°C colder than that of Turbet et al. (2016), who use an installation 74.3 $Wm^{-2}$ larger. Our dynamic ocean simulation with the Turbet et al. (2016) installation value (*Control-High*) has a dayside maximum only 3°C warmer than *Control* with small differences in the pattern of surface temperature and sea ice (Figs. 1c, 2c), which illustrates the damping effect of a dynamic ocean on the climate response to external forcing. Larger differences occur on the nightside: Our minimum temperature decreases by 30°C without ocean heat transport, and increases by 8°C when installation is increased, but in neither case do we approach the very cold nightside temperatures (-123°C) simulated by Boutle et al. (2017).

*b.) Effect of elevated greenhouse gas concentrations*

Our *Archean* experiments are not surprisingly warmer than *Control* (Table 3), but the progression in surface temperature as greenhouse gases increase from *Low* to *Med* to *High* is not monotonic except for the nightside minimum. The added dayside surface warming from $CO_2$ and $CH_4$ is weak (only 2-3°C warmer peak temperatures). In other respects, the *Archean* simulations differ from our others in surprising ways. For example, despite being globally warmer, *Archean Med* is actually colder at the equator away from the substellar region (Fig. 1d). Its sea ice cover is slightly greater than *Control* (Fig. 2d) and much greater than *Archean Low* (Table 3) and it thus has a smaller habitable fraction ($f_{hab}$ = .38) despite higher greenhouse gas concentrations. *Archean Med* has somewhat more ice-free area than *Control* at high latitudes, but lacks the equatorial band of open ocean water on the nightside.

This seemingly counter-intuitive behavior appears to be due to competition between the radiative effect of increasing greenhouse gases and changes in the atmospheric dynamical response to installation from a cool star as the composition changes. Table 4 summarizes some of the relevant aspects of this competition. Installation is identical in *Control* and the *Archean* simulations, but atmospheric absorption of the mostly near-infrared flux of Proxima Centauri by the increased $H_2O$ (Fig. 4a), $CO_2$, and $CH_4$ in the *Archean* simulations increases with each increase in $CO_2$ and $CH_4$. Thus, progressively less SW is absorbed at the surface at the dayside peak. Combined with the added LW greenhouse effect of these gases, the result is an increasingly statically stable substellar temperature profile from *Control* to *Archean Low* to *Archean Med* to *Archean High*. Figure 4b shows this for *Control* vs. *Archean Med*.

Several previous GCM studies have explored the effect of planet rotation on the atmospheric dynamics of synchronously rotating rocky planets. Noda et al. (2017) identified a transition at ~20 d rotation period between a dynamical regime dominated by a day-night circulation with weak zonal winds at longer periods and a regime characterized by a Matsuno-Gill Rossby-Kelvin wave response to the substellar heating and equatorial superrotating winds at somewhat shorter periods. Edson et al. (2011) found a similar transition at a rotation period of ~4-5 d. Neither study is directly comparable to ours because of various differences in the models; for example, neither assumes a dynamic ocean or a cool star spectrum. Nonetheless, the regime transition seen in both studies may be relevant to the differences between our *Control* and *Archean* Proxima b climates.

*Control* is clearly in the Rossby-Kelvin regime. The upper troposphere has a strong wavenumber 1 standing oscillation (Fig. 5a) with the anticyclone centered near/downwind of the substellar region and cyclones at high latitudes on the nightside (see Figs. 6b,c of Noda et al., 2017). The NW-SE tilt of the flow in the northern hemisphere and SW-NE tilt in the southern hemisphere downstream of the substellar point create a strong equatorward flux of zonal momentum by eddies (Fig. 5c), similar to that seen in simulations of slowly rotating hot Jupiters by Showman et al. (2015, their Fig. 7a). This produces strong (~60 m s$^{-1}$) nightside equatorial superrotation (Fig. 5e, Table 4), as predicted by Showman and Polvani (2010, 2011). Between the evening terminator and antistellar point fairly strong subsidence (not shown), presumably the downward branch of the Kelvin wave pattern, advects potentially warm air from the dayside downward and extends the superrotation to the surface (Fig.5e), helping to maintain above-freezing equatorial ocean temperatures there.

The *Archean* simulations also appear to be in the Rossby-Kelvin regime, but barely so; Figure 5 shows analogous fields for *Archean Med*. Relative to *Control*, *Archean Med* exhibits only a weak stationary wave (Fig. 5b), much weaker equatorward eddy momentum transports (Fig. 5d), and consequently weaker (~30 m s$^{-1}$) superrotation (Fig. 5f, Table 4). Near the surface, *Archean Med* exhibits a pure day-night circulation more like that of the almost non-rotating regime of Noda et al. (2017) (see their Fig. 5a) and much weaker nightside subsidence. Thus, there is less adiabatic warming of descending air and no transport of dayside near-surface air to the nightside. As a result the equatorial ocean surface remains frozen on the nightside and into the morning on the dayside, increasing sea ice cover. *Archean Low* and *Archean High* show similarly weak eddy momentum fluxes and equatorial superrotation (Table 3).

Edson et al. (2011) argue that the regime transition from Rossby-Kelvin to day-night occurs as the Rossby radius of deformation (which is related to the spatial scale of the waves) approaches the radius of the planet. For equatorial waves, the deformation radius is $L_d = (NH/\beta)^{1/2}$, where $N = [(g/\Theta)\partial\Theta/\partial z]^{1/2}$ is the Brunt-Väisälä frequency, $g$ is gravity, $\Theta$ is potential temperature, $z$ is altitude, $H$ is the scale height, $\beta = 2\Omega/a$ is the meridional gradient of the Coriolis parameter at the equator, $\Omega$ is the rotational angular frequency of the planet, and $a$ is the planet radius. In both Edson et al.'s (2011) and Noda et al.'s (2017) experiments, the transition from the Rossby-Kelvin to the day-night regime is accomplished by increasing the rotation period (thus increasing $L_d$ by decreasing $\beta$). In our synchronous rotation simulations, the rotation period is fixed. $L_d$ changes instead because the temperature and lapse rate change as greenhouse gases increase (Fig. 4b), as seen in other studies of M star planets (Shields et al., 2013; Rugheimer et al., 2015). For Proxima Centauri b, $a = 7127.335$ km in our simulations. We estimate that $L_d < a$ for *Control* and $L_d > a$ for all the *Archean* simulations, albeit by modest amounts. Furthermore, for *Archean Med* and *Archean High*, although the simulations are in equilibrium when averaged over sufficiently long time scales, significant variability remains on time scales of ~3-8 Earth years, with global surface temperature fluctuating by up to ~5°C and sea ice fraction by up to ~15% (absolute). In addition, we find that small changes in model physics such as removing the spectral dependence of sea ice albedo can cause the *Archean Med* climate and circulation to change significantly. This supports our impression that the *Archean* simulations are close to the regime transition and may exhibit bistable behavior, as Edson et al. (2011) found near the transition point.

In the lower stratosphere, the *Archean* simulations are tens of degrees warmer than *Control*, and their weaker cold traps allow stratospheric $H_2O$ to build up to levels two orders of magnitude greater than in our other simulations (Table 3). This is still ~10 times smaller than the traditional moist greenhouse water loss limit (Kasting et al, 1993), but then the $CO_2$ abundances assumed for these experiments are not excessive, considering that the silicate-carbonate cycle may not operate efficiently on an aquaplanet (Abbot et al., 2012). Thus, regardless of an early super-luminous phase or XUV irradiation of the planet from its star, an aquaplanet Proxima Centauri b with elevated $CO_2$ may experience significant water loss by the traditional diffusion-limited escape mechanism.

*c.) Fresh and salty ocean planets*

A wide range of Proxima b interior compositions are consistent with the observational constraint on its mass, including a significant water mass underlain by high-pressure ices and a negligible water mass (relative to planet mass) with a partly silicate mantle below (Brugger et al., 2016). Given our lack of knowledge of how much water was originally delivered to Proxima b, the nature of its source, and the history of its escape, an ocean-covered Proxima b could have a wide range of salinities. As discussed by Cullum et al. (2016), this has two implications for climate: (1) Changes in the density-driven thermohaline component of the ocean circulation and thus ocean heat transport, and (2) changes in the freezing point depression of sea water and thus the temperature threshold for surface liquid water.

Our experiments *Zero Salinity* (S = 0 psu) and *High Salinity* (S = 230 psu), patterned loosely after Cullum et al. (2016), test this effect. For S = 0, the freezing point of seawater is 0°C (as opposed to -1.9°C in *Control,* for which S = 35.4), while for S = 230 psu seawater freezes at -19°C, several degrees warmer than the eutectic point temperature. The ocean model in ROCKE-3D uses an equation of state from Fotonoff and Millard (1983) that is intended for salinities appropriate to Earth's ocean and linearly extrapolates the density of seawater for S > 40 psu, including the pressure dependence of the thermal expansion and haline contraction coefficients. At our coldest realized upper ocean temperature of -12.6°C (at which the local S = 232.8 psu) in *High Salinity*, this equation of state predicts a seawater density ~1230 kg m$^{-3}$. By comparison, Sharqawy et al. (2010) present a polynomial fit for seawater density at atmospheric pressure valid to within 0.1% up to S = 160 psu and down to 0°C; extrapolating this fit to 232.8 psu and -12.6°C gives a density of 1192 kg m$^{-3}$, 3% less than the prediction from ROCKE-3D. Our primary interest, though, is the effect of the depression of the freezing point rather than the details of the ocean circulation.

Figure 6 shows surface temperature and sea ice cover for the *Zero Salinity* and *High Salinity* cases. Compared with *Control*, the spatial patterns of warm temperatures and open ocean are very different: In *Zero Salinity*, the "lobster" has no tail, producing a less asymmetric pattern about the substellar longitude (Figs. 6a,b). In *High Salinity,* broad regions of open ocean extend into the extratropics, with only a weak remnant of the lobster pattern. More important for habitability is that the peak dayside temperature for *Zero Salinity (High Salinity)* is 7°C warmer (3°C colder) than in *Control*, although the global mean temperatures are colder (warmer), respectively. The effects on sea ice cover and thus fractional planet habitability are more dramatic, with 10% less ice-free surface area in *Zero Salinity* ($f_{hab}$ = .32) and an almost ice-free planet ($f_{hab}$ = .87) in *High Salinity* despite surface temperatures on the latter planet that never exceed 0°C. The difference in remotely detectable equilibrium temperature between these two extreme planet scenarios, though, is only 6°C (Table 3).

In all three simulations, surface downwelling occurs in a narrow band near and upstream of the substellar point where the wind-driven circulation associated with the Rossby gyres converges; upwelling is concentrated in a narrow equatorial band east of the substellar region. In *Zero Salinity*, salinity forcing of the density-driven circulation (due to regional imbalances between surface evaporation E and precipitation P) is absent, and peak surface ocean current velocities are about half those in *Control.* The primary density-driven circulation is downwelling near the sea ice margins, due to the unique property of fresh water that its density peaks at 4°C. The equatorial westerly current seen in *Control* is almost absent in *Zero Salinity* except in the substellar region; most horizontal ocean heat transport takes place in the wind-driven subtropical gyres. Other than this, though, there is little evidence of surface salinity forcing of the circulation in *Control.* For example, in the convergence regions of the Rossby-Kelvin pattern off the equator where P > E, there is only a weak ocean upwelling anomaly.

In *High Salinity,* nearly ice-free conditions are due primarily to the freezing point depression of the very salty water. The Rossby-Kelvin wave surface wind circulation still exists and produces surface ocean current convergence and downwelling of the ocean circulation at the equator, but the primary westerly surface ocean current extends to midlatitudes upstream of the substellar point. This is associated with an equator-pole density gradient that occurs upstream of the substellar region near the ocean surface but becomes almost zonal at depth, with cold, saltier water in the polar regions.

Figure 7 shows the ocean vertical structure along the equator for the three experiments with different salinity. In *Zero Salinity*, stratification is due to temperature differences alone (Fig. 7a), so the potential density structure (Fig. 7b) is identical to that of potential temperature. For *Control* and *High Salinity*, potential density is controlled mostly by salinity differences, so we show the salinity structure instead. In *Zero Salinity* the ocean near and downstream of the substellar point is warm and stably stratified (potential density increasing downward), more weakly stratified on the nightside, and almost neutrally stratified upstream of the substellar point (Figs. 7a,b). In *Control*, warm water is more concentrated near the ocean surface downstream of the substellar point (Fig. 7c), with vertically uniform salinity there (Fig. 7d).  Near both terminators, salty water collects near the bottom, producing very stable stratification except near the antistellar point. The ocean structure in *High Salinity* is similar to that of *Control* but with the warm dayside water even closer to the surface (Fig. 7e) and very salty water at the ocean bottom at almost all longitudes.

*d.) 3:2 spin-orbit resonance aquaplanets*

As pointed out by Ribas et al. (2016), Proxima b may be in a 3:2 spin-orbit resonance rather than in synchronous rotation.  We conducted two aquaplanet experiments to explore the habitability of such states. The first simulation (*3:2e0)* assumes eccentricity e = 0. This is done strictly for comparison to Turbet et al. (2016), since a 3:2 resonance state requires nonzero eccentricity (Goldreich and Peale, 1966; Ribas et al., 2016). The second simulation (*3:2e30)* assumes e = 0.30, between the most likely estimate of e = 0.25 suggested by Brown (2017) and the upper limit of e = 0.35 suggested by Anglada-Escudé et al. (2016).

Figure 8 shows incident SW radiation at the top of the atmosphere, surface temperature, and sea ice cover for these simulations.  For e = 0, all longitudes receive the same time-mean instellation (Fig. 8a). The resulting atmospheric circulation is similar to that of a moderately slowly rotating Earth (Del Genio and Suozzo, 1987), with a Hadley cell in each hemisphere that extends most of the way to the pole, rather than the day-night circulation of synchronously rotating planets. This produces upper level westerly jets in midlatitudes and easterly trade winds at low latitudes in the near-surface Hadley cell return flow, but a very weak westerly equatorial surface ocean current.

In 3:2 resonance, for which the rotation period is 7.5 d, the stellar day is 22.4 d, much shorter than the several Earth year thermal response time of the 900 m ocean. This prevents thick sea ice from building up at low latitudes where instellation is greatest. But the thin (12 m) first ocean layer is able to partially freeze at night, with ~20-30% sea ice cover (Fig. 8e) of 0.2-0.3 m thickness and otherwise open ocean with sea surface temperature at the freezing point near the equator in the time mean (although the surface air temperature is several degrees colder, Fig. 8c). This planet is the closest to being uninhabitable ($f_{hab}$ = .21) at the surface of all the scenarios considered in this paper. Our result is fairly similar to that reported by Turbet et al. (2016), but theirs is achieved with an instellation 74 W m$^{-2}$ higher than ours. Boutle et al. (2017) obtain an equatorial waterbelt using the same instellation we use, but only when they assume a dark surface albedo; with a brighter albedo, their model produces transient open ocean conditions.

Of more interest is what occurs with non-zero eccentricity, which is much more likely for an asynchronously rotating planet. In this case the stellar heating pattern has two maxima on opposite sides of the planet and weak heating at longitudes in between (Fig. 8b), as shown by Dobrovolskis (2007) and Brown et al. (2014). In the Boutle et al. (2017) model this leads to "eyeball" liquid water regions on either side of the planet and ice in between. In our simulations, though, the tropical ocean remains above freezing with no sea ice even at weakly irradiated longitudes that receive only 133 W m$^{-2}$ instellation in the time mean (Fig. 8d). This climate is reminiscent of past "slushball" periods on Earth (Sohl et al., 2017). The Proxima b tropical waterbelt however is zonally asymmetric (Fig. 8d), retaining some memory of the preferentially heated longitudes. Peak open ocean occurs slightly downstream of peak instellation due to modest eastward ocean heat transport at the most strongly heated longitudes. This planet is somewhat more habitable ($f_{hab}$ = .51) than the synchronously rotating *Control* ($f_{hab}$ = .42). For both asynchronous planets, $T_{max}$ and $T_{min}$ in Table 3 refer to equatorial and polar temperatures rather than dayside and nightside conditions.

*e.) Earth-like land-ocean planets*

Moderate $CO_2$ concentrations such as those assumed in most of our simulations may be more likely if some land is exposed to stabilize the climate via the carbonate-silicate cycle (Abbot et al., 2012). We conducted two experiments with synchronous rotation but an Earth land-ocean distribution, differing only in substellar longitude. Experiment *Day-Ocean* places the substellar point at the dateline (180°), so that the strongest heating is over the Pacific Ocean. *Day-Land* has the substellar point at 30°E, over the African and Eurasian continents.

Figures 9 and 10 show several aspects of the climates of these two planets, and Table 3 summarizes their global mean properties. In many respects these two simulations most closely resemble our *Thermo* experiment, which ignores ocean heat transport. Warm temperatures and surface liquid water (Figs. 9a,b and 10a,b) are restricted to the dayside in a pattern not too different from the eyeball Earth configuration of *Thermo*. The difference is that for *Thermo*, the eyeball pattern is produced artificially by excluding a process (ocean heat transport) that operates in nature. In *Day-Ocean* and *Day-Land*, it is due to the continental configuration of Earth, which is dominated by two large land masses oriented primarily north-south. This produces partly enclosed ocean basins that limit ocean heat transport to the nightside, except in the narrow midlatitude Southern Ocean.

The *Day-Ocean* and *Day-Land* climates and circulations bear some resemblance to terrestrial winter and summer monsoons, respectively. In *Day-Ocean*, the near-surface flow (Fig. 9c) is generally directed offshore toward the central Pacific instellation peak, where convergence (and heavy precipitation, not shown) occurs in several bands determined partly by the influence of the Rossby-Kelvin wave pattern and partly by details of the continental distribution. In *Day-Land,* the flow is primarily onshore toward the African continent (Fig. 10c), where convergence and heavy precipitation occur. Subsurface liquid water reservoir levels (Fig. 10d, volume liquid soil water per volume of soil), an indicator of land surface habitability, are highest in *Day-Land* in a band where the near-surface convergence occurs. Areas with > 0.3 volumetric fraction (0.4855 $m^3$ $m^{-3}$ is the saturation value for our assumed soil mix and depth) are comparable to the wettest, most heavily vegetated places on Earth. This is surrounded by a broader dayside continental region where subsurface water volumetric fractions are 0.10-0.15, typical of annual means for semi-arid to arid terrestrial climates. In *Day-Ocean*, the only habitable land areas are in coastal regions near the terminators (Fig. 9d) that are moderately wet, despite being far from the mid-Pacific convergence zone (Fig. 9c).

Elsewhere in *Day-Ocean* and *Day-Land*, land is snow-covered. After 15,000 orbits (460 Earth years), net radiation, surface temperature, planetary albedo, and snow depth have reached equilibrium. Snow depth nowhere exceeds 5 m water equivalent. However, lakes form on the nightside, and although these contain only ~$10^{-4}$ as much water as the accumulated snow, they are continuing to grow, in effect serving as a proxy land ice model. Yang et al. (2014) estimate from an offline land ice model that for a planet similar to ours, ice sheets of O(1 km) thickness should form on the nightside, though dayside water would remain for an inventory > ~10% of Earth's oceans. Thus, we might expect the ocean in *Day-Ocean* and *Day-Land* to partly but not completely migrate to the nightside if ROCKE-3D could run to complete water cycle equilibrium (which would require tens of thousands of years or more, well beyond the capabilities of GCMs).

**4. DISCUSSION AND CONCLUSIONS**

We have shown that with a dynamic ocean, a hypothetical ocean-covered Proxima Centauri b with an atmosphere similar to modern Earth's can have a habitable climate with a broad region of open ocean, extending to the nightside at low latitudes.  This is true for both synchronous rotation and a 3:2 spin-orbit resonance with moderate eccentricity.  Because of its weak instellation, however, Proxima Centauri b's climate is unlikely to resemble modern Earth's. "Slushball" episodes in Earth's distant past with cold but above-freezing tropical oceans (Sohl et al, 2017) are better analogs.  The extent of open ocean depends on the salinity assumed.  Elevated greenhouse gas concentrations produce some additional warming, but this is limited for any M-star planet by the reduced penetration of near-infrared starlight to the surface, and for Proxima b in particular, by its existence near a possible dynamical regime transition.  "Eyeball Earth" scenarios with a warm substellar region do not occur in reality because of ocean heat transport, unless exposed land that creates a partially enclosed dayside ocean basin is present.  We consider some further implications of our results below.

*a.) Implications for the habitable zone concept*

The *nominal* equilibrium temperature reported by exoplanet observers (e.g., by Anglada-Escudé et al., 2016 for Proxima b) and used to define the habitable zone is based on installation and an assumed Earth-like planetary albedo of 0.3. This is an unreliable indicator of habitability, since albedo and eccentricity vary among planets (Barnes et al., 2015), and planetary albedo varies with stellar spectral type (Kopparapu et al., 2013). The nominal equilibrium temperature (228 K) is identical in all our experiments except *Control-High*, yet mean (maximum) surface temperatures vary among the experiments by 27 (34) °C and the ice/snow-free fraction of the surface varies from .20-.87, largely because planetary albedo varies from .16-.28 (Table 3). None of our simulations reach the 0.3 Earth value of planetary albedo, because of absorption of the enhanced near-IR incident starlight by greenhouse gases and the reduced Rayleigh scattering of the atmosphere, as found in previous studies of M-star planets (Kasting et al., 1993; Kopparapu et al., 2013; Shields et al., 2013; Rugheimer et al., 2015).

The *actual* planetary equilibrium temperature (determined by the *absorbed* stellar flux), on the other hand, is a fairly good predictor of surface habitability in most of our simulations, even though it is determined by emission primarily from the mid-troposphere – there is a 0.72 correlation between $T_{eq}$ and $f_{hab}$ and a .96 correlation between $T_{eq}$ and $T_{mean}$ in Table 3. This is derived from a very limited set of experiments, all having identical atmospheric pressures and many having identical compositions, but it suggests that observations of rocky planet thermal phase curves may be of at least limited use for a subset of planets for which other information can rule out very un-Earthlike conditions (e.g., strongly irradiated runaway hothouse planets like modern Venus).

A linear fit to the results in Table 3 yields the prediction $f_{hab}$ = *.0418 $T_{eq}$ − 9.5089.* Out-of-sample tests that exclude each experiment in turn show that the resulting fits generally predict the actual $f_{hab}$ to within 0.1 or better. The biggest outlier is the most un-Earthlike planet, *High Salinity*, for which the out-of-sample regression predicts $f_{hab}$ .51, much less than the actual .87. This is not surprising given that the physical process that keeps much of this experiment's surface liquid is non-radiative and thus is not captured by what is basically an energy budget metric. The second biggest disagreement occurs for *3:2e0* (predicted $f_{hab}$ = .43 vs. actual $f_{hab}$ = .21), the only experiment in which instellation is invariant with longitude. Thus, different interpretations of phase curves may be necessary for planets in different spin-orbit configurations.

Our *High Salinity* planet does not fit the traditional notion of habitability, since its time mean surface temperature never exceeds 0°C anywhere. Nonetheless, bodies of water on Earth such as the Dead Sea with conditions near 230 psu salinity are extensively inhabited by halophilic life (e.g., Fendrihan et al., 2006). Although these examples are at warm temperatures, Mykytczuk et al. (2012, 2013) report that the bacterium *Planococcus halocryophilus* strain Or1 grows and divides in 180 psu water at temperatures as low as -15°C, not far from the ocean conditions in *High Salinity*, and is metabolically active down to -25°C.

The lower temperature limit of life is not yet defined (De Maayer et al., 2014). Standard cryopreservation temperatures for microbial cultures can be as low as -196°C in a cryoprotectant to maintain viable if not active cultures (Kirsop and Doyle, 1991). In nature, psychrophilic (or cryophilic) organisms have been observed to support metabolic functions in a hypersaline Antarctic spring as low as -20 °C at 220-260 psu (Lamarche-Gagnon et al., 2015). All of the domains of life (Archaea, Bacteria, and Eucaryo) include halotolerant, halophilic, and haloextremophiles able to live or requiring environments with salinities ranging through freshwater/seawater interfaces, typical seawater at 35 psu, up to 359 psu at NaCl saturation, from extreme cold to hot temperatures (Oren, 2002).

Thus, if we think more broadly about what constitutes a habitable planet (Cullum et al., 2016), it is reasonable to imagine a cold but inhabited Proxima Centauri b with a very salty shallow remnant of an earlier extensive ocean in which halophilic bacteria are the dominant life form, especially if the ocean is in contact with a silicate seafloor (Glein and Shock, 2010). We note that Jupiter's moon Europa, an object of interest in the search for life in our Solar System, may have an ocean nearly saturated with salt (Hand and Chyba, 2007). Objects with subsurface oceans such as Europa and Saturn's moon Enceladus have been considered irrelevant to the search for life in other stellar systems because of the challenge of remotely detecting subsurface life, but they may be a useful analog for water-rich, weakly irradiated tidally locked exoplanets that maintain a small region of substellar or equatorial liquid water that may be detectable.

An aquaplanet cannot support a carbonate-silicate cycle to stabilize climate, because the seafloor will likely be insufficiently exposed to the temperature changes needed for weathering feedbacks (Abbot et al., 2012). Also, the $CO_2$ cycle is subject to an unstable feedback due to temperature-dependent ocean dissolution of $CO_2$ (Kitzmann et al., 2015; Levi et al., 2017). 3-D processes might counteract such tendencies if increased solubility in cold ocean regions offsets the effect of decreased solubility in the substellar region, with recycling following transport of colder ocean water back to the substellar region. Simulating the pattern of $CO_2$ transport in a coupled ocean-atmosphere model and identifying the installation and stellar temperature required for a stable cycle would help constrain long-term prospects for life on planets such as Proxima b.

*b.) The role of clouds*

Terrestrial GCMs are traditionally categorized by their climate sensitivity, i.e., the change in surface temperature per unit change in external radiative forcing (e.g., a change in greenhouse gas concentrations or insolation). For most models, the sensitivity lies in the range ~0.5-1.5 °C m$^2$ W$^{-1}$ (Andrews et al, 2012).

Following Boutle et al. (2017), we estimate our Proxima b climate sensitivity by comparing *Control* and *Control-High*. The latter has 18.6 Wm$^{-2}$ more instellation averaged over the planet and its mean surface temperature is 7.5°C warmer, yielding a climate sensitivity of 0.40 °Cm$^2$W$^{-1}$. This is greater than the ~0.27 °Cm$^2$W$^{-1}$ sensitivity found by Boutle et al. (2017) for a model with the same radiation scheme but different parameterizations of clouds, convection, and surface albedo. Like their model, though, our Proxima b sensitivity is less than our parent Earth GCM's sensitivity of ~0.6 °Cm$^2$W$^{-1}$ (Schmidt et al., 2014), although that sensitivity is in response to greenhouse gas rather than solar forcing.

Boutle et al. (2017) suggest that low sensitivity occurs for a synchronously rotating planet because low-level clouds occur primarily on the nightside and do not affect reflected sunlight. Low clouds generally decrease in response to warming in Earth climate change simulations (Zelinka et al., 2016), decreasing planetary albedo and creating a positive feedback. A simple albeit imperfect measure of the effect of clouds on climate is the cloud radiative forcing (CRF), defined as the difference in shortwave (SW) and longwave (LW) radiative flux with vs. without clouds, the latter calculated offline in parallel with the actual model integration. Positive *CRF* means that the planet is warmer with clouds than without them. Thus usually *SWCRF* < 0 because clouds are brighter than most planet surfaces, while *LWCRF* > 0 because most clouds have a greenhouse effect.

In *Control*, the planet is nearly overcast (Fig. 11a). The nightside has almost ubiquitous low cloud (Fig. 11b), similar to Boutle et al. (2017), and has moderate to large middle (Fig. 11c) and high (Fig. 11d) cloud cover as well, but these are optically thin (vertically integrated liquid + ice water contents < 40 g m$^{-2}$ everywhere and < 5 g m$^{-2}$ in most locations). On the dayside, however, optically thick clouds (integrated ice + liquid water contents of 100-300 g m$^{-2}$) generated by moist convection and large-scale upwelling in response to the direct stellar heating occur, as predicted by Yang et al. (2013) and other GCM studies for synchronously rotating planets. The dayside has > 90% cloud fraction everywhere except for a small region near the low latitude morning terminator, where cloudiness is closer to 80%. Small clear to partly cloudy regions do occur occasionally on time scales of an orbit or more, but primarily near the terminators and at high latitudes. If this is typical of synchronously rotating exoplanets with extensive surface oceans, detection of surface biosignatures or even the ocean itself will be difficult. Analysis of the stochastic properties of these clouds would be needed to quantify the detectability of such features from different viewing angles and phases as may be observed by future direct imaging missions.

Figure 12 shows the response of these clouds to the instellation increase from 881.7 W m$^{-2}$ in *Control* to 956 W m$^{-2}$ in *Control-High*. Increased instellation produces regional increases and decreases in low cloud fraction (Fig. 12a) but no systematic change on the dayside or nightside. High cloud fraction increases weakly in the substellar region and to a greater extent in the equatorial region on the nightside (Fig. 12b). More important is the dayside strengthening (i.e., becoming more negative) of SWCRF (Fig. 12e), especially downstream of the substellar point, which appears to be due partly to increasing cloud optical thickness associated with an increase in cloud water path, much but not all of it in high ice clouds but some in the liquid phase in lower altitude clouds as well (Figs. 12c,d). This results in a more negative *SWCRF* there, i.e., a negative SW cloud feedback over the dark substellar ocean. Significant increases in *LWCRF* also occur on both the dayside and nightside over open ocean; this may slightly underestimate the true LW cloud feedback since the cloud forcing approach conflates the effects of cloud changes with collocated non-cloud feedbacks (Soden et al., 2004).

Table 5 shows global mean and dayside maximum values of *SWCRF* and *LWCRF*, as well as the global mean clear-sky greenhouse effect $G_a = <\sigma T_{surf}^4> - <\sigma T_{eq}^4> - <LWCRF>$ (where the brackets indicate that the fluxes are area-averaged), for each simulation. In general, the warmer the peak substellar temperature $T_{max}$ or the larger the day-night difference $T_{max} - T_{min}$, the stronger (more negative) *SWCRF* is, but this only explains part of the variance in *SWCRF* (correlation coefficients -0.74 in each case).

*SWCRF* and *LWCRF* are only weakly correlated globally (-0.28), but their maximum values in the substellar convection region are highly correlated (-0.89), as they are in tropical deep convective regions on Earth (Kiehl, 1994). Unlike Earth, however, where Kiehl (1994) found that *SWCRF* and *LWCRF* nearly offset each other ($R = -LWCRF/SWCRF \sim 1$) to produce near-zero net cloud forcing in convective regions, substellar *LWCRF* is considerably smaller than *SWCRF* in our Proxima b simulations ($R$ = 0.16-0.59).

This appears to occur for several reasons: the day-night circulation in our synchronously rotating cases, the height of convective cloud systems, and non-condensing greenhouse gas concentrations. Unlike Earth's tropics, where thick high clouds are common but variable in time due to propagating disturbances and not strongly correlated with time of day, on a synchronously rotating planet such clouds are usually present and coincident with strong instellation. Consequently, peak SWCRF exceeds that on Earth. Second, the Proxima b convective clouds are not as deep as Earth's, because of the weak instellation relative to what Earth receives from the Sun and the stronger near-infrared absorption by $H_2O$, $CO_2$, and $CH_4$ that stabilizes our Proxima b troposphere (e.g., Fig. 4). Especially low values ($R$ = 0.16-0.19) occur for the three *Archean* simulations, which have among the weakest peak *LWCRF* (Table 5) due to the elevated greenhouse gas concentrations that reduce the opacity contrast between clear and cloudy sky.

Our experiments support the nightside "radiator fin" analogy to Earth's subtropics discussed by Yang and Abbot (2014). The best predictor of nightside minimum surface temperature in our simulations is the clear-sky greenhouse effect $G_a$ (correlation 0.93), which is much larger than *LWCRF* (Table 5). $G_a$ in turn is weakest for experiments with reduced day-night heat and water vapor transport (*Thermo, Zero Salinity, Day-Ocean, Day-Land*), all of which are quite cold on the nightside, as opposed to *Control, Control-High,* and *High Salinity*, which have high $G_a$ values and more moderate nightside climates. The warmest nightside conditions other than *High Salinity* are for the *Archean* simulations, a reflection of those experiments' elevated $CO_2$ and $CH_4$ levels.

*c.) Relevance to other known exoplanets*

Our simulations are specific to best-estimate parameters for Proxima Centauri b and assume an atmospheric composition and water inventory that are at present unknown, but the results should be at least qualitatively applicable to similar recently discovered potentially rocky planets orbiting other very cool M stars. For example, the TRAPPIST-1 system of planets (Gillon et al., 2017) orbits a star of temperature 2650 K, about 400 K colder than Proxima Centauri. Similar concerns about water loss exist for this system as for Proxima Centauri b (Bolmont et al., 2017), but given the unknown evolutionary path of the system planets that retain water cannot be ruled out. One planet (TRAPPIST-1 e) is 0.918 times the radius of Earth, receives an incident stellar flux 0.662 times the flux incident on Earth, very close to that received by Proxima b, and has a synchronous rotation period of 6.1 d, slightly more than half that of Proxima b. A simulation of TRAPPIST-1 e by Wolf (2017) with a thermodynamic ocean aquaplanet and a 1 bar $N_2$ atmosphere with a modest $CO_2$ concentration yields a cold but partly habitable "eyeball" planet with mean surface temperature similar to our *Thermo* simulation. Thus, our *Control* simulation might be a useful analog for this planet if it has an Earthlike atmosphere.

Astudillo-Defru et al. (2017) have announced 9 new nearby planets observed by the HARPS spectrograph. One of these (GJ 273 b) orbits a 3382 K M star, receives 1.06 times the solar flux received by Earth, is 2.89 or more times the mass of Earth, and has a rotation period of 18.6 d. If the actual mass is close to the minimum mass and/or the density is greater than Earth's, GJ 273 b could be a rocky planet. If so, the instellation and rotation values are close enough to those of our simulations for us to anticipate a GJ 273 b climate somewhat warmer than (and thus having more habitable surface area than) but otherwise similar to our Proxima b *Control-High* case given the same type of atmosphere. [For example, Fujii et al. (2017) simulate substellar temperatures of ~290-295 K for an Earth-size planet with a static ocean orbiting the stars GJ 876 and Kepler-186, which are somewhat cooler and warmer, respectively, than GJ 273.] Another planet, GJ 3293 d, orbits a 3480 K star, receives 0.59 times the solar flux incident on Earth, has a minimum mass 7.6 times that of Earth, and rotates once every 48.1 d; depending on its actual mass and density, it may or may not be a rocky planet. If it is, its climate could be similar to those described here for Proxima b given a similar atmosphere, but it is more likely to be in the slowly rotating day-night dynamical regime than the Rossby-Kelvin regime.

Most recently, Dittman et al. (2017) have reported a planet orbiting a 3131 K M star, LHS 1140 b, that is 1.4 times the radius and 6.6 times the mass of Earth and receives 0.46 times the instellation that Earth receives. This level of stellar heating makes it marginal for habitability. Fujii et al. (2017), using a GCM with a static ocean, find that for GJ 876, a star similar in temperature to LHS 1140, an Earth-size planet with a 1 bar $N_2$ atmosphere and a minimal amount of $CO_2$ (1 ppmv) has a substellar surface temperature of ~262 K (and is thus ice-covered) for instellation 0.4 times Earth's; the temperature is ~281 K for instellation 0.6 times Earth's. This suggests that with larger greenhouse gas concentrations more like our *Archean High* case and a dynamic ocean, scenarios may exist in which LHS 1140 b can be habitable, especially if any ocean is saltier than Earth's.

The population of potentially habitable rocky exoplanets in M star systems has now suddenly reached the point at which it will soon be possible to assess the demographics of this class of planet. Future research will benefit from the context provided by climate model studies of this class as a whole in anticipation of efforts to begin detecting and characterizing their atmospheres. Several such general GCM studies have already occurred (Bolmont et al., 2016; Kopparapu et al., 2016; Fujii et al. 2017; Wolf et al., 2017), but exploration of the parameter space of factors that influence the climates of these planets has only just begun.

**ACKNOWLEDGEMENTS.** We are grateful to Andrew Lincowski, Giada Arney, Eddie Schwieterman, Jacob Lustig-Yeager, and Vikki Meadows of VPL, who created the Proxima Centauri stellar spectrum and generously shared it with us for this paper. We also thank Bill Kovari for assistance with figures. This work was supported by the NASA Astrobiology Program through collaborations arising from our participation in the Nexus for Exoplanet System Science, and by the NASA Planetary Atmospheres Program. Computing resources for this work were provided by the NASA High-End Computing (HEC) Program through the NASA Center for Climate Simulation (NCCS) at Goddard Space Flight Center.

**AUTHOR DISCLOSURE STATEMENT:** No competing financial interests exist.

**TABLE 1.** Wavelength ranges (µm) for simulations using 29 SW and 12 LW bands.

| SW | | | | LW | |
|---|---|---|---|---|---|
| Band # | Range | Band # | Range | Band # | Range |
| 1 | 0.200-0.385 | 16 | 1.56-1.62 | 1 | 25-10000 |
| 2 | 0.385-0.500 | 17 | 1.62-1.68 | 2 | 18.18-25 |
| 3 | 0.500-0.690 | 18 | 1.68-1.80 | 3 | 12.5-13.33 and 16.95-18.18 |
| 4 | 0.690-0.870 | 19 | 1.80-1.94 | 4 | 13.33-16.95 |
| 5 | 0.870-0.900 | 20 | 1.94-2.00 | 5 | 10.01-12.5 |
| 6 | 0.900-1.08 | 21 | 2.00-2.14 | 6 | 8.93-10.01 |
| 7 | 1.08-1.12 | 22 | 2.14-2.50 | 7 | 8.33-8.93 |
| 8 | 1.12-1.16 | 23 | 2.50-2.65 | 8 | 7.63-8.33 |
| 9 | 1.16-1.20 | 24 | 2.65-2.85 | 9 | 7.09-7.63 |
| 10 | 1.20-1.30 | 25 | 2.85-3.15 | 10 | 4.57-7.09 |
| 11 | 1.30-1.34 | 26 | 3.15-3.60 | 11 | 4.15-4.57 |
| 12 | 1.34-1.42 | 27 | 3.60-4.10 | 12 | 3.34-4.15 |
| 13 | 1.42-1.46 | 28 | 4.10-4.60 | | |
| 14 | 1.46-1.52 | 29 | 4.60-20.0 | | |
| 15 | 1.52-1.56 | | | | |

**TABLE 2.** Description of GCM simulations. $S_o$ is the stellar constant; S is salinity; e is the eccentricity. Simulations with an * use 29/12 SW/LW radiation bands; all others use 6/9 SW/LW bands.

| # | Name | Description |
|---|------|-------------|
| 1 | Control | 0.984 bar, $N_2$ + 376 ppmv $CO_2$ atmosphere, dynamic ocean, aquaplanet, synchronous rotation, $S_o$ = 881.7 $Wm^{-2}$, S = 35.4 psu |
| 2 | Thermo | Like Control but with a thermodynamic ocean |
| 3 | Control-High | Like Control but with $S_o$ = 956 $Wm^{-2}$ |
| 4 | Archean Low* | Like Control but 638 ppmv $CO_2$, 450 ppmv $CH_4$ |
| 5 | Archean Med* | Like Control but 900 ppmv $CO_2$, 900 ppmv $CH_4$ |
| 6 | Archean High* | Like Control but 10000 ppmv $CO_2$, 2000 ppmv $CH_4$ |
| 7 | Zero Salinity | Like Control but S = 0 psu |
| 8 | High Salinity | Like Control but S = 230 psu |
| 9 | 3:2e0 | Like Control but in 3:2 resonance with e=0 |
| 10 | 3:2e30 | Like Control but in 3:2 resonance with e=0.30 |
| 11 | Day-Ocean | Like Control but with Earth land-ocean distribution and substellar point over Pacific |
| 12 | Day-Land | Like Day-Ocean but with substellar point over Africa |

**TABLE 3.** Climate variables for each experiment: Maximum, minimum and mean surface temperature ($T_{surf}$), fraction of surface not covered by sea ice or snow ($f_{hab}$), equilibrium temperature ($T_{eq}$), planetary albedo ($A_p$), 100 hPa maximum water vapor volume mixing ratio ($H_2O$). Ocean freezing points are 0°C for S = 0 psu, -1.9°C for S = 35.4 psu, and -19°C for S = 230 psu.

| # | Experiment | $T_{surf}$ (°C) | | | $f_{hab}$ | $T_{eq}$ (K) | $A_p$ | $H_2O$ (ppmv) |
|---|---|---|---|---|---|---|---|---|
| | | Max | Min | Mean | | | | |
| 1 | Control | 3 | -59 | -21 | .42 | 237 | .234 | 1.1 |
| 2 | Thermo | 19 | -89 | -37 | .20 | 233 | .282 | 2.0 |
| 3 | Control-High | 6 | -51 | -13 | .55 | 242 | .232 | 1.4 |
| 4 | Archean Low | 6 | -48 | -15 | .50 | 240 | .183 | 108.2 |
| 5 | Archean Med | 5 | -46 | -18 | .38 | 239 | .181 | 178.1 |
| 6 | Archean High | 6 | -42 | -10 | .56 | 240 | .161 | 439.8 |
| 7 | Zero Salinity | 10 | -69 | -28 | .32 | 235 | .254 | 1.0 |
| 8 | High Salinity | 0 | -43 | -10 | .87 | 241 | .178 | 0.5 |
| 9 | 3:2e0 | -3 | -62 | -20 | .21 | 237 | .204 | 0.4 |
| 10 | 3:2e30 | 3 | -53 | -13 | .51 | 241 | .201 | 0.6 |
| 11 | Day-Ocean | 18 | -77 | -28 | .44 | 235 | .253 | 1.2 |
| 12 | Day-Land | 31 | -92 | -33 | .37 | 234 | .253 | 1.3 |

**TABLE 4.** Parameters related to the dynamical regime shift from the *Control* simulation to the three *Archean* simulations.

| Experiment | Clear sky incident SW at surface (W m$^{-2}$) | Rossby radius of deformation (km) | Equatorial jet maximum zonal wind speed (m s$^{-1}$) |
|---|---|---|---|
| Control | 640 | 6446 | 58 |
| Archean Low | 484 | 7973 | 28 |
| Archean Med | 472 | 8122 | 31 |
| Archean High | 409 | 8232 | 34 |

**TABLE 5.** Maximum and global mean SWCRF and LWCRF, and global mean clear-sky atmospheric greenhouse effect ($G_a$), all in units of W m$^{-2}$, for each experiment.

| # | Experiment | SWCRF$_{max}$ | <SWCRF> | LWCRF$_{max}$ | <LWCRF> | $G_a$ |
|---|---|---|---|---|---|---|
| 1 | Control | -191 | -23 | 61 | 12 | 52 |
| 2 | Thermo | -317 | -33 | 125 | 11 | 22 |
| 3 | Control-High | -195 | -29 | 58 | 13 | 63 |
| 4 | Archean Low | -183 | -21 | 35 | 7 | 66 |
| 5 | Archean Med | -120 | -21 | 19 | 5 | 58 |
| 6 | Archean High | -128 | -19 | 20 | 6 | 84 |
| 7 | Zero Salinity | -206 | -30 | 72 | 11 | 36 |
| 8 | High Salinity | -200 | -19 | 75 | 16 | 72 |
| 9 | 3:2e0 | -54 | -15 | 32 | 10 | 54 |
| 10 | 3:2e30 | -80 | -19 | 39 | 12 | 66 |
| 11 | Day-Ocean | -212 | -32 | 80 | 13 | 39 |
| 12 | Day-Land | -278 | -28 | 122 | 11 | 32 |

**FIGURE LEGENDS**

**FIG. 1.** Surface temperature distributions for **(a)** *Control*, **(b)** *Thermo*, **(c)** *Control-High*, and **(d)** *Archean Med*. The substellar point is at the center of each figure. The color bar range is defined in this and other surface temperature figures so that the transition from yellow to blue occurs at the freezing point of seawater for the salinity chosen.

**FIG. 2.** As in Figure 1 but for sea ice cover.

**FIG. 3.** *Control* experiment distributions of **(a)** surface air wind velocity, **(b)** surface ocean current velocity, **(c)** sea ice mass flux, and **(d)** sea ice thickness. The black region in panel (d) is complete open ocean with no sea ice.

**FIG. 4.** Substellar point vertical profiles of **(a)** specific humidity (kg $H_2O$ kg air$^{-1}$) and **(b)** temperature (°C) for the *Control* and *Archean Med* simulations.

**FIG. 5. (a,b)** Geopotential height at 335 hPa (substellar longitude at center)**, (c,d)** zonally averaged northward eddy momentum flux, and **(e,f)** pressure vs. longitude distribution of equatorial zonal wind (antistellar point at center) for (left) *Control* and (right) *Archean Med*.

**FIG. 6. (a,c)** Surface temperature and **(b,d)** sea ice cover for the *Zero Salinity* (upper panels) and *High Salinity* (lower panels) simulations.

**FIG. 7.** Depth-longitude profiles of **(a,c,e)** ocean potential temperature for *Zero Salinity*, *Control*, and *High Salinity*; **(b)** ocean potential density for *Zero Salinity*; **(d,f)** ocean salinity for *High Salinity*.

**FIG. 8.** **(a,b)** Incident SW radiation at the top of the atmosphere, **(c,d)** surface temperature, and **(e,f)** sea ice cover for the (left) *3:2e0* and (right) *3:2e30* simulations.

**FIG. 9. (a)** Surface temperature, **(b)** snow and sea ice cover, **(c)** surface wind velocity, and **(d)** subsurface land liquid water for experiment *Day-Ocean.*

**FIG. 10.** As in Figure 9 but for experiment *Day-Land.*

**FIG. 11.** *Control* **(a)** total, **(b)** low**, (c)** middle, and **(d)** high cloud cover.

**FIG. 12.** Changes in **(a)** low cloud fraction, **(b)** high cloud fraction, **(c)** total cloud water path (liquid + ice), **(d)** cloud ice water path, **(e)** SWCRF, and **(f)** LWCRF in response to a change in instellation (*Control-High – Control*).

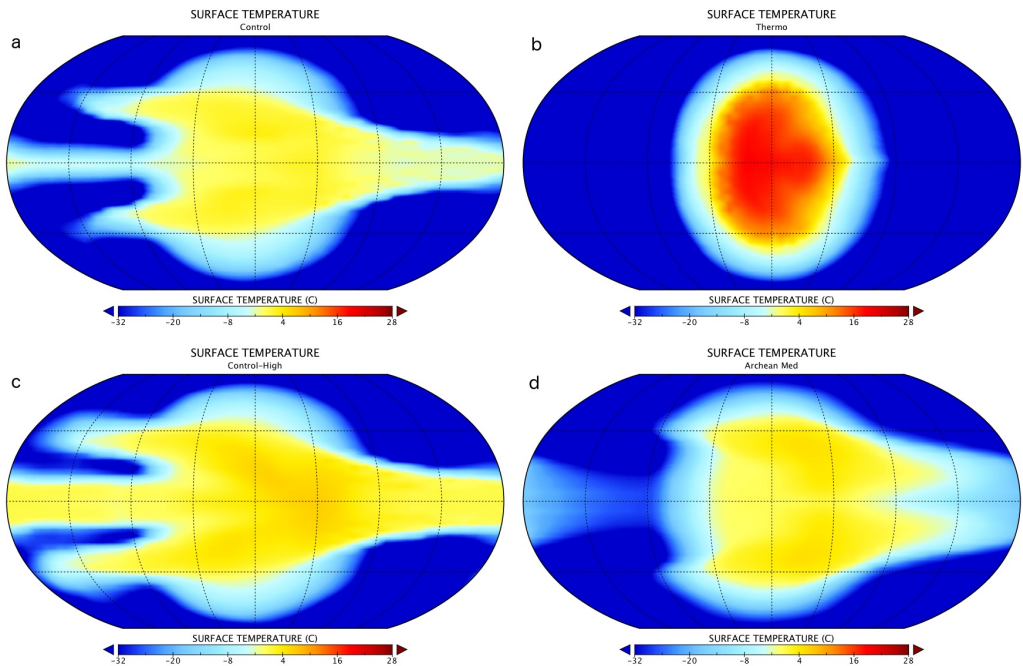

**FIG. 1.** Surface temperature distributions for **(a)** *Control*, **(b)** *Thermo*, **(c)** *Control-High*, and **(d)** *Archean Med.* The substellar point is at the center of each figure. The color bar range is defined in this and other surface temperature figures so that the transition from yellow to blue occurs at the freezing point of seawater for the salinity chosen.

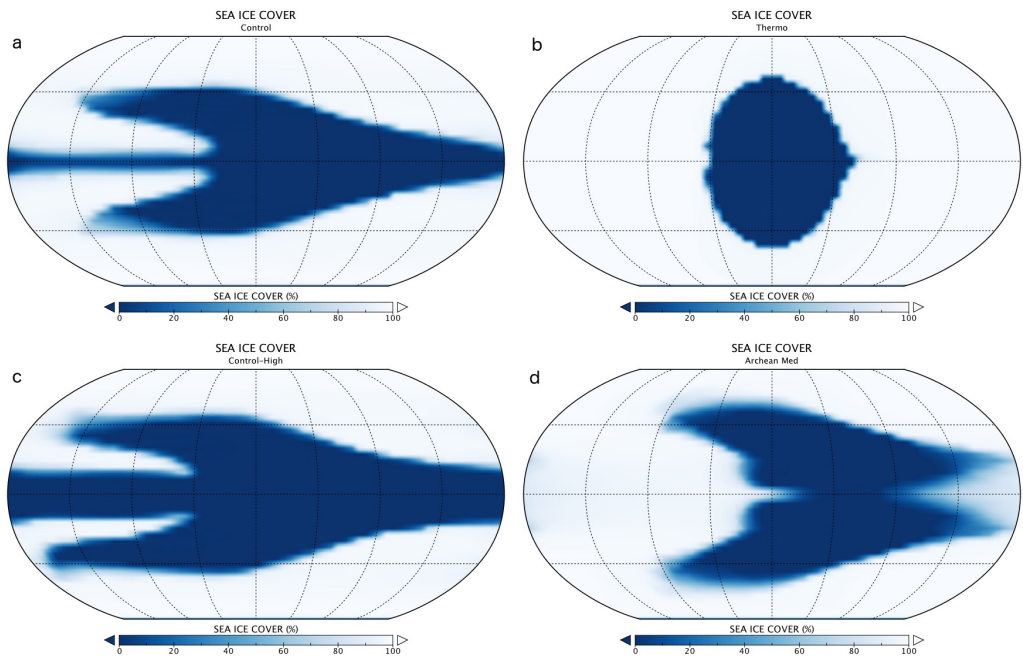

**FIG. 2.** As in Figure 1 but for sea ice cover.

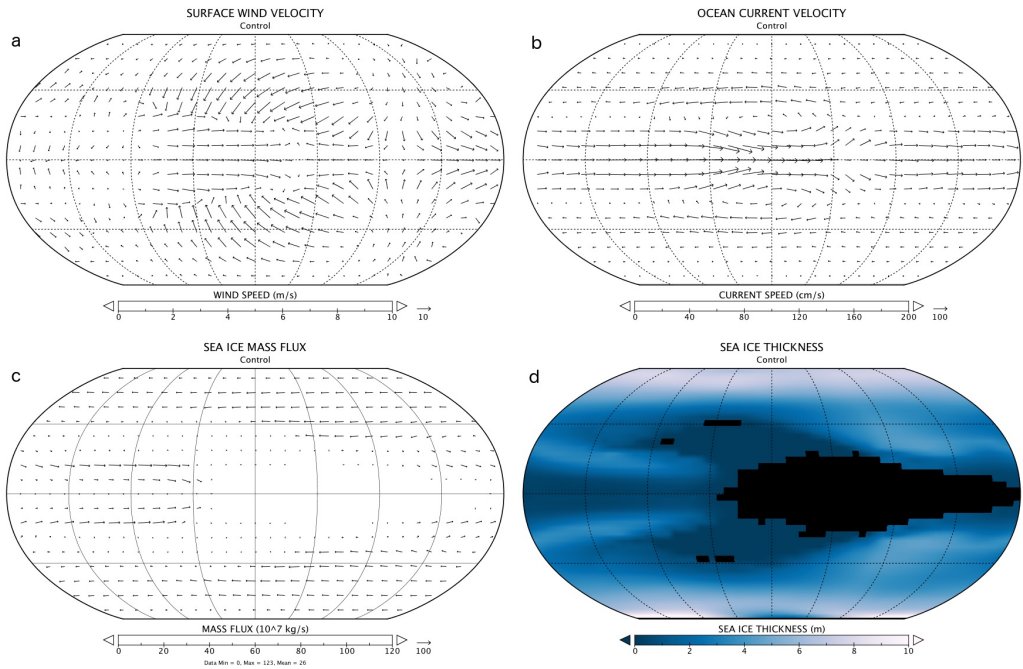

**FIG. 3.** *Control* experiment distributions of **(a)** surface air wind velocity, **(b)** surface ocean current velocity, **(c)** sea ice mass flux, and **(d)** sea ice thickness. The black region in panel (d) is complete open ocean with no sea ice.

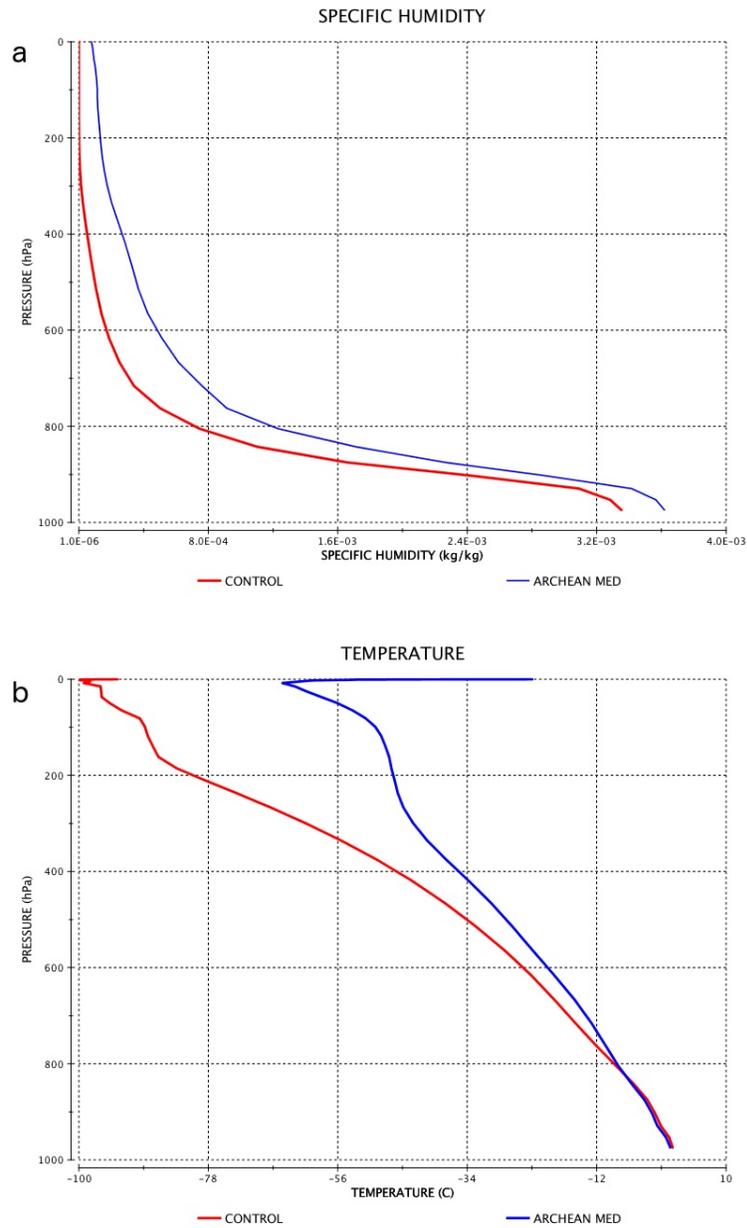

**FIG. 4.** Substellar point vertical profiles of **(a)** specific humidity (kg H$_2$O kg air$^{-1}$) and **(b)** temperature (°C) for the *Control* and *Archean Med* simulations.

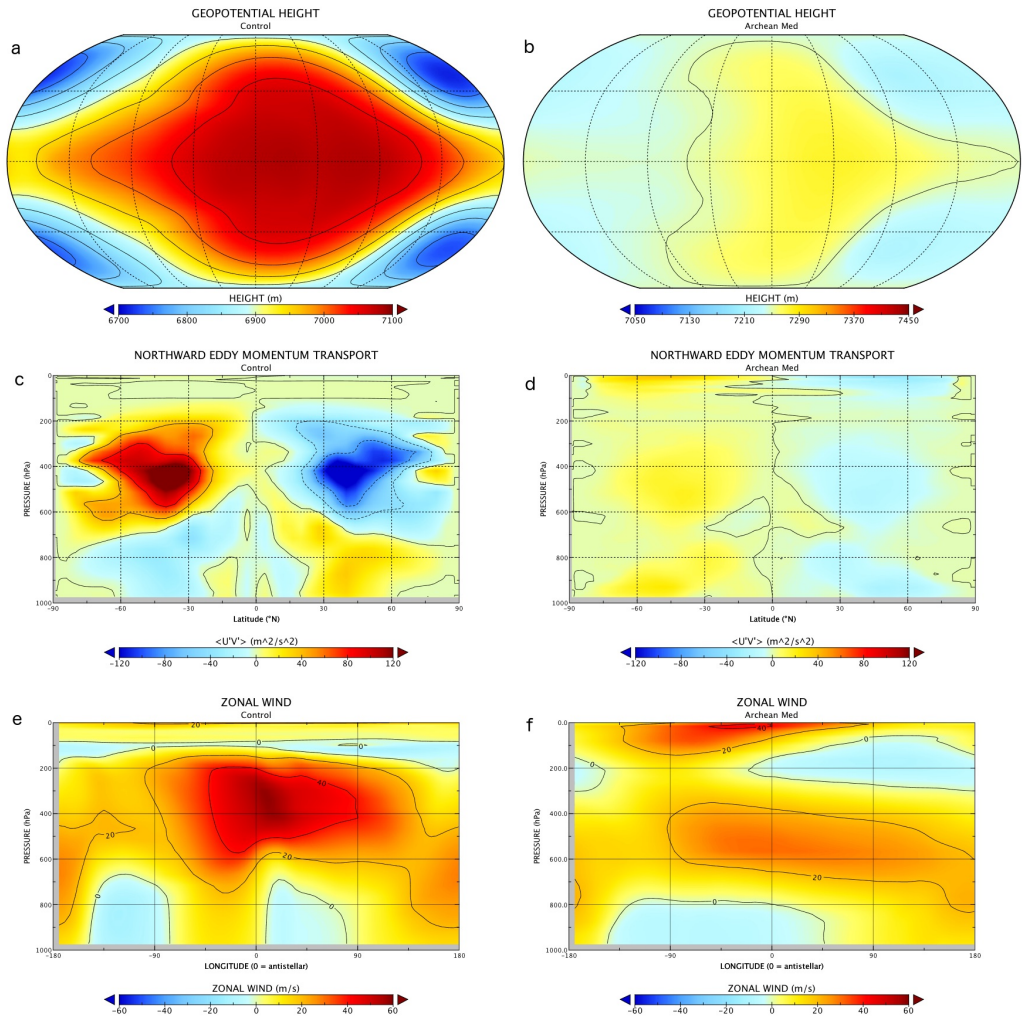

**FIG. 5. (a,b)** Geopotential height at 335 hPa (substellar longitude at center)**, (c,d)** zonally averaged northward eddy momentum flux, and **(e,f)** pressure vs. longitude distribution of equatorial zonal wind (antistellar point at center) for (left) *Control* and (right) *Archean Med.*

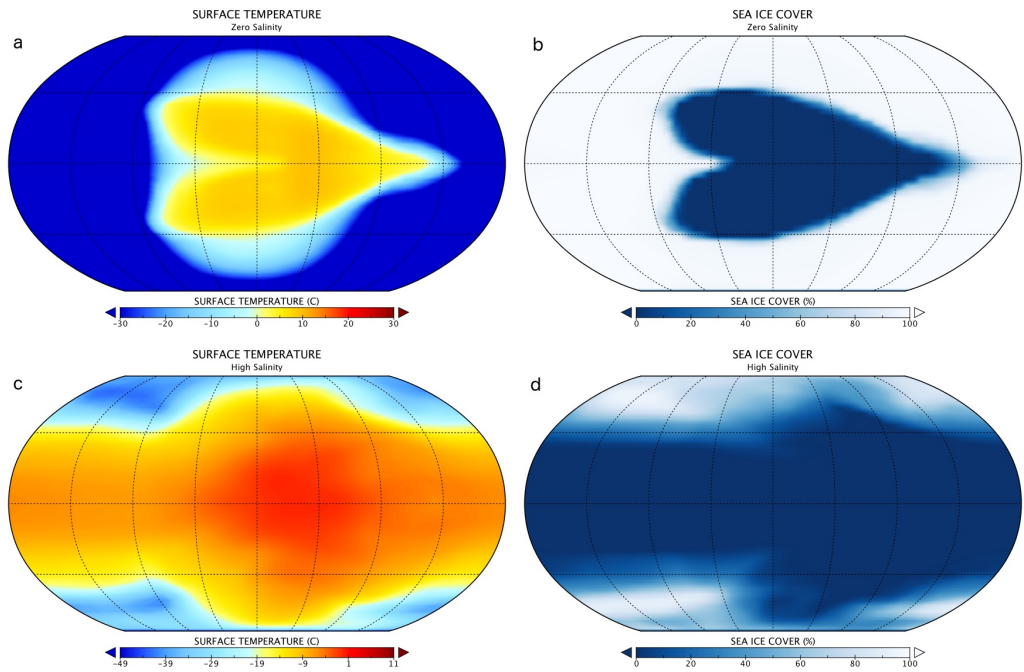

**FIG. 6.** **(a,c)** Surface temperature and **(b,d)** sea ice cover for the *Zero Salinity*
(upper panels) and *High Salinity* (lower panels) simulations.

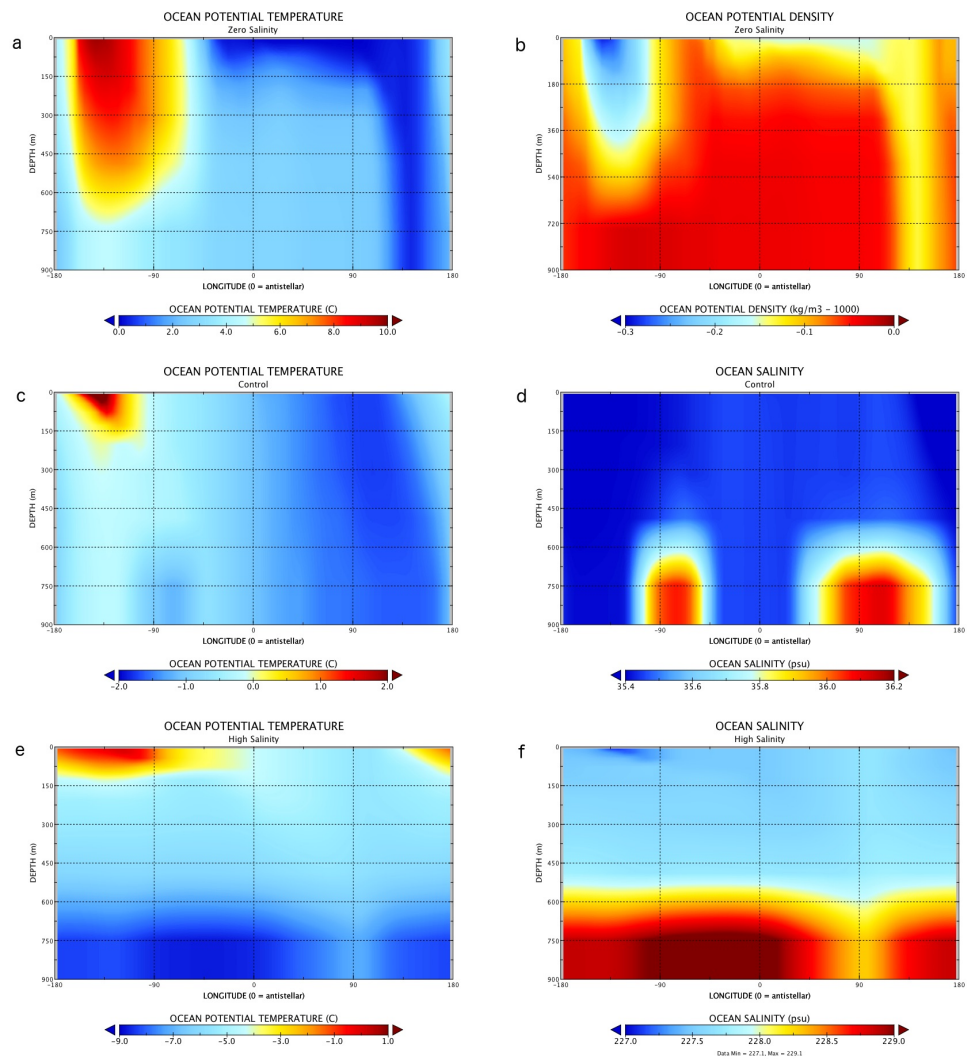

**FIG. 7.** Depth-longitude profiles of **(a,c,e)** ocean potential temperature for *Zero Salinity*, *Control*, and *High Salinity*; **(b)** ocean potential density for *Zero Salinity*; **(d,f)** ocean salinity for *Control* and *High Salinity*.

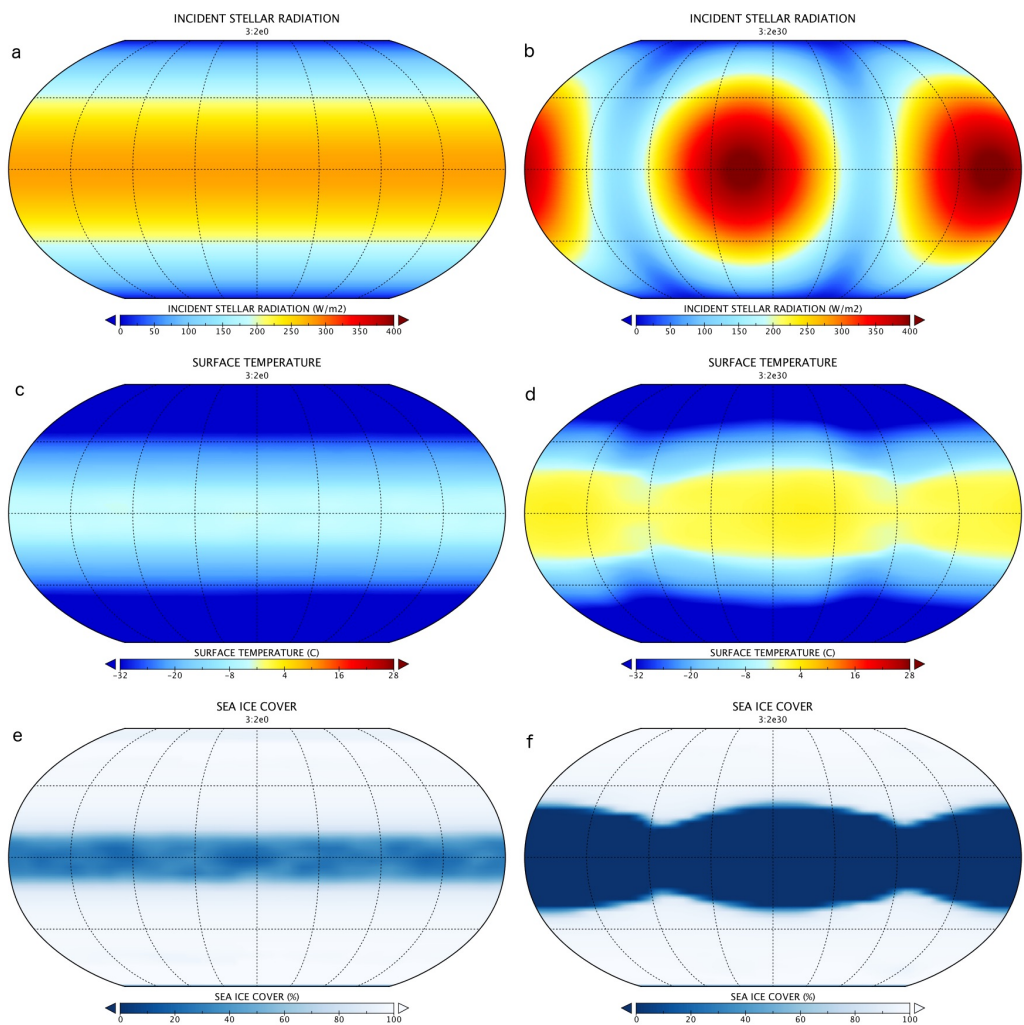

**FIG. 8. (a,b)** Incident SW radiation at the top of the atmosphere, **(c,d)** surface temperature, and **(e,f)** sea ice cover for the (left) *3:2e0* and (right) *3:2e30* simulations.

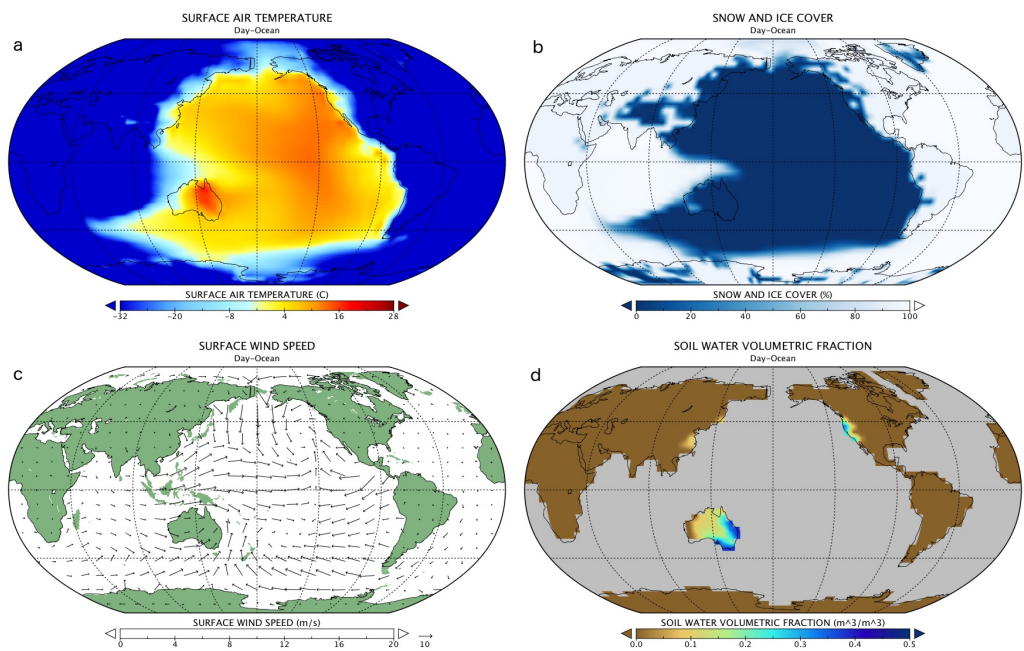

**FIG. 9. (a)** Surface temperature, **(b)** snow and sea ice cover, **(c)** surface wind velocity, and **(d)** subsurface land liquid water for experiment *Day-Ocean.*

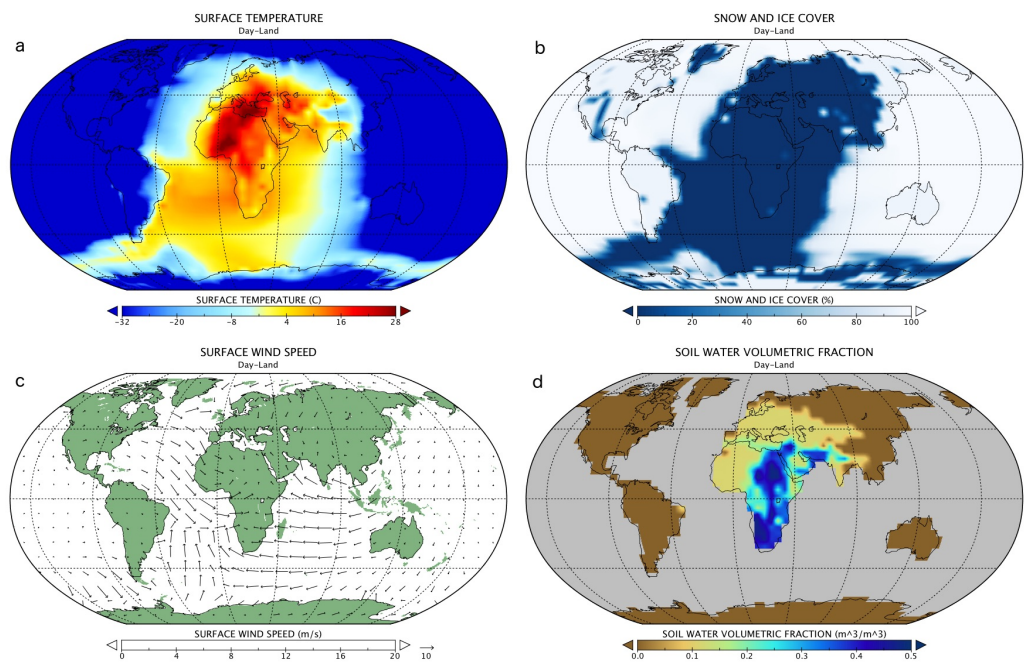

**FIG. 10.** As in Figure 9 but for experiment *Day-Land.*

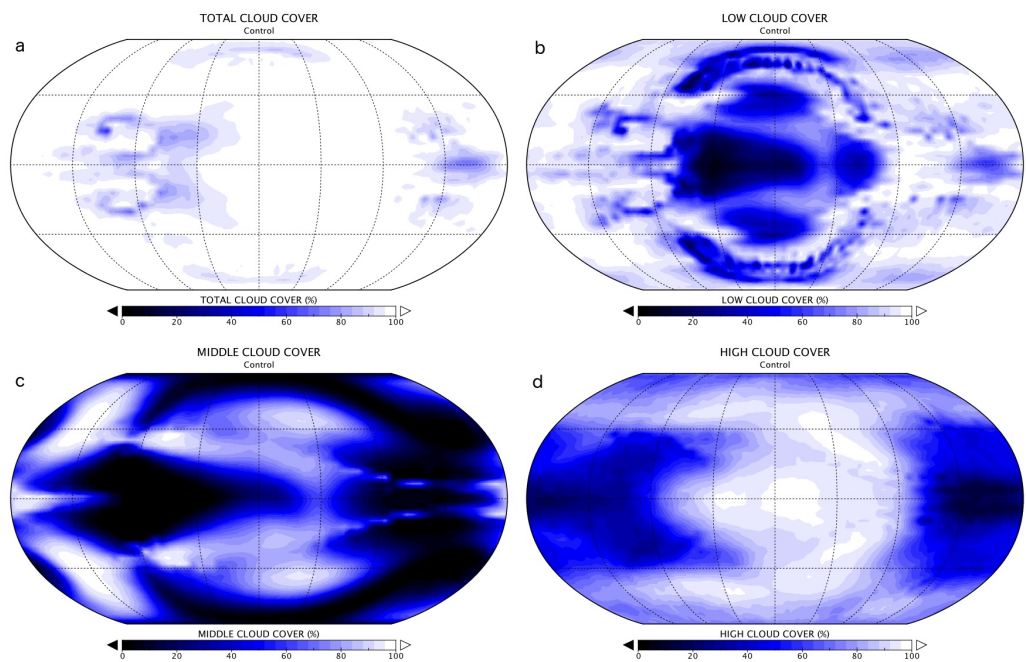

**FIG. 11.** *Control* **(a)** total, **(b)** low**, (c)** middle, and **(d)** high cloud cover.

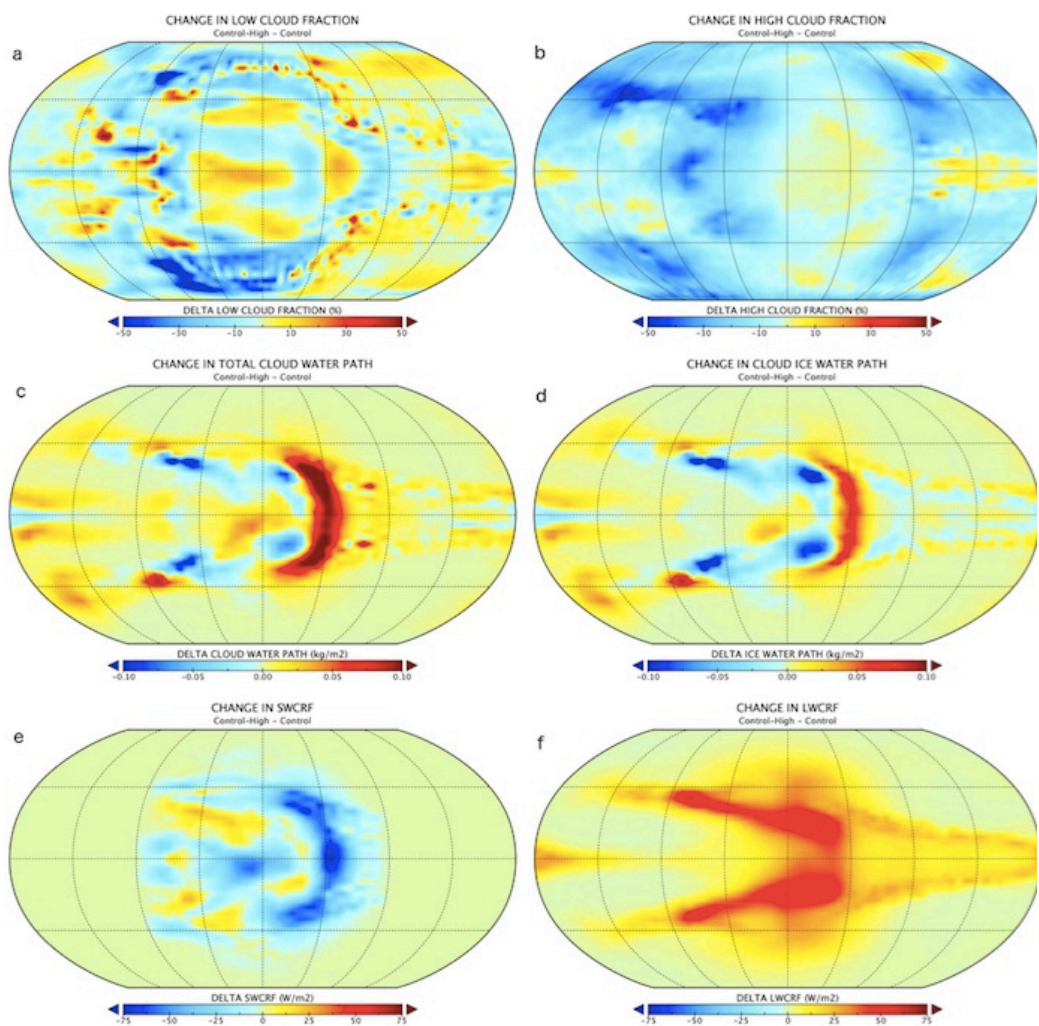

**FIG. 12.** Changes in **(a)** low cloud fraction, **(b)** high cloud fraction, **(c)** total cloud water path (liquid + ice), **(d)** cloud ice water path, **(e)** SWCRF, and **(f)** LWCRF in response to a change in instellation (*Control-High – Control*).